\documentstyle[12pt]{article}

\def\di{\displaystyle}
\def\&{&\di}
\def\bg{\begin{eqnarray}\begin{array}{rcl}\displaystyle}
\def\eg{\end{array} &\di    &\di   \end{eqnarray}}
\def\bm#1{\begin{eqnarray}\begin{array}{#1}\di}
\def\bmo#1{\begin{eqnarray*}\begin{array}{#1}\di}
\def\bgo{\begin{eqnarray*}\begin{array}{rcl}\displaystyle}
\def\ego{\end{array} &\di    &\di \nonumber  \end{eqnarray*}}

\def\btensor#1#2{\renew\left#1\begin{array}{#2}\di}
\def\brtensor#1#2#3{\ren#3\left#1\begin{array}{#2}}
\def\botensor#1#2{\renew\left#1\begin{array}{#2}}
\def\etensor#1{\end{array}\right#1}

\def\eq#1{(\ref{#1})}

\def\d{d}

\def\tr{\mbox{tr}}
\def\Tr{\mbox{Tr}}

\def\ov{\over}
\def\p{\partial\llap{/}}

\def\dr{{D\!\llap{/}\,}}
\def\Dr{{D\llap{/}}}

\def\r{R^\psi}

\def\N{\mbox{l}\!\mbox{N}}

\def\CD{{\cal D}}

\def\CF{{\cal F}}

\def\CM{{\cal M}}
\def\CN{{\cal N}}
\def\CO{{\cal O}}
\def\CP{{\cal P}}

\def\CT{{\cal T}}
\def\CU{{\cal U}}

\date{\today}

\def\ren#1{\renewcommand{\arraystretch}{#1}}
\def\rene{\renewcommand{\arraystretch}{1.9}}
\def\renew{\renewcommand{\arraystretch}{1}}
\rene

\newcommand{\mysection}[1]{\section{#1}\setcounter{figure}{0}
\setcounter{table}{0}\setcounter{equation}{0}}

\begin{document}

\begin{titlepage}

\parindent=12pt
\baselineskip=20pt
\textwidth 15 truecm
\vsize=23 truecm
\hoffset=0.7 truecm

\begin{flushright}
   FSUJ-TPI-05/96 \\
  hep-th/9605037     
      \end{flushright}
\par
\vskip .5 truecm
\large \centerline{\bf Exact Flow Equations and the $U(1)$-Problem} 
\par
\vskip 1 truecm
\normalsize
\begin{center}
{\bf J.~M.~Pawlowski}\footnote{current address: School of 
Theoretical Physics, Dublin Institute for Advanced Studies, 10
  Burlington Road, Dublin 4, Ireland\\ e-mail: jmp@stp.dias.ie}\\
\it{Theor.-Phys. Institut, Universit\"at Jena\\ 
Fr\"obelstieg 1\\
D-07743 Jena\\
Germany}
\end{center}
 \par
\vskip 2 truecm
\normalsize
\begin{abstract} 
The effective action of a $SU(N)$-gauge theory coupled to fermions 
is evaluated at a large infrared cut-off scale $k$ within the path integral
approach. The gauge field measure includes topologically non-trivial
configurations (instantons). Due to the explicit infrared
regularisation there are no gauge field zero modes. The Dirac operator
of instanton configurations shows a zero mode even
after the infrared regularisation, which leads to 
$U_A(1)$-violating terms in
the effective action. These terms are calculated in the limit of large
scales $k$.
\end{abstract}
       
\vfill
       
 PACS: 11.10.Gh, 11.15.-q, 11.30.Er\\ 
Keywords: Instanton-induced effects, Effective Actions, Flow Equations 
\end{titlepage}


\mysection{Introduction}
One of the outstanding problems in QCD is the $U(1)$-problem. This
problem can be reformulated as the question about the missing ninth
light pseudo-scalar meson, the $\eta'$. There have been several
attempts to solve this problem at least qualitatively (see reviews 
\cite{christoshuryak},\cite{thooft1} and references therein). 
However, a quantitative description is still missing
and discrepancies between the different approaches have not been 
clarified 
yet. This is not surprising as a quantitative description must 
include non-perturbative QCD-effects. 
                      
 One approach  to tackle this intricate problem is to
use flow equations (or exact renormalisation group equations
\cite{Wilson}) in continuum QFT. They were exploited by Polchinski
and subsequently by many other authors to simplify proofs of
perturbative renormalisability \cite{Polchinski}. More recently
they have been 
investigated also as a powerful tool to study non-perturbative 
physics. The
flow equation describes the scale dependence of the effective action
$\Gamma_k$, where $k$ is an infrared cut-off scale. The starting point is an
effective action $\Gamma_{k_0}$ which depends only on modes 
with momenta larger than $k_0$. Then the flow
equation is used to integrate out successively the momenta 
smaller than the cut-off scale $k_0$. 
           
This method has been applied also to gauge theories 
\cite{becchi}-\cite{stis}. Here
one must employ a gauge
invariant effective action. An appropriate way to deal with gauge 
invariance is to use the background field formalism as presented in 
\cite{abbott}. Within this
formalism it is possible to define an infrared
regularised version of the effective action, which is gauge invariant
\cite{reutwett}. However, the background field dependence of the
cut-off term leads to additional terms in the Slavnov-Taylor 
identities (STI). 
A second possibility was proposed in
\cite{bonetal},\cite{ellwanger}, where the
infrared cut-off term explicitly
breaks gauge invariance. 
In both approaches one derives so-called modified
STIs. It has been 
shown, that the original STI are restored in the
limit, where the cut-off is removed ($k\rightarrow 0$). Moreover, the
modified STI are compatible with the flow equations. Thus, given 
an effective action fulfilling the modified STI at the starting scale,
the integrated effective action fulfills the modified STI at arbitrary
scales \cite{ellwanger,stis}.  
    
In using flow equations for explicit calculations one has 
to deal with the following
problems: In principle the flow equation consists of an 
infinite system 
of coupled differential equations. In most interesting cases it is not 
possible to solve these equations analytically and one has to
truncate the system. Thus the problem is how to find truncations 
valid in the regime of interest and to formulate general validity
checks for these truncations. Furthermore one needs to know the 
effective action at the starting point $k_0$. Even for large scales 
$k_0$ the effective action consists of an infinite number of terms. 
In principle it should be possible to start at
 a large initial scale $k_0$ with the (finite) number of 
terms which depend on relevant couplings. 
The corrections due to the terms neglected at $k_0$ remain of the 
order $1/k_0$ for all scales $k$ because of universality. 
But this argument is only valid if one is dealing
with the non-truncated flow equation. Since one has to 
truncate the system of differential equations it is important to
calculate the effective action at the starting scale $k_0$ as accurately
as possible.

To be more explicit, we study the behaviour of the flow
equation under global $U_A(1)$-variations. For that purpose let us 
briefly outline the derivation of the flow equation: In the path integral 
formulation we 
achieve a smooth infrared cut-off by adding a
 scale-dependent term to the action 
(e.g.\ \cite{reutwett,ellwanger})
\bm{c}\label{cut-off1} 
\Delta_k S[\Phi] = \frac{1}{2} \int \Phi
R^\Phi_k[P_\Phi]\Phi,
\eg 
where $P_\Phi^{-1}$ is proportional to 
the bare propagator of $\Phi$ and $\Phi$ is a shorthand notation for all 
fields. The regulator $R_k^\Phi$ has the following properties: 
\bm{ccccc} \label{limits}
R_k^\Phi[x]\stackrel{{x\ov k^2}\rightarrow 0}{\longrightarrow} 
k^{4-2 d_\Phi} \frac{x}{|x|}, & &\di 
R_k^\Phi[x]
\stackrel{{x\ov k^2}\rightarrow \infty}{\longrightarrow}  0,
\eg 
where $d_\Phi$ are the dimensions of the fields $\Phi$. 
Hence the 
cut-off term \eq{cut-off1} 
effectively suppresses modes with momenta  
$p^2\ll k^2$ in the generating functional. 
For modes with large momenta $p^2\gg k$ the cut-off term 
vanishes and in this regime the theory remains unchanged. In the limit 
$k\rightarrow 0$ we approach the full generating functional since 
the cut-off term is removed. An infinitesimal 
variation of the generating functional with respect to
 $k$ is described by the flow equation. 
For the generating functional of 1PI Green functions, 
the effective action $\Gamma_k$, the flow equation 
can be written in the form (see for example \cite{reutwett,ellwanger})
\bg  \label{flow}
\partial_t\Gamma_k[\Phi]=  \frac{1}{2}\Tr\left\{
\partial_t R^\Phi_k[P_\Phi]
(\Gamma^{(2)}_k[\Phi]+R_k^\Phi[P_\Phi])^{-1}\right\},
\eg 
where $t=\ln k$ and 
the trace $\Tr$ denotes 
a sum over momenta, indices and the different fields $\Phi$. 
$\Gamma^{(2)}_k$ is the second derivative of 
$\Gamma_k$ with respect to the fields $\Phi$. Note that $\partial_t R_k^\Phi$ 
serves as a smeared-out $\delta$-function in momentum space peaked about 
$p^2\approx k^2$. Thus by varying the scale $k$ towards smaller $k$ 
according to \eq{flow} one successively integrates out 
momentum degrees of freedom related to $p^2\approx k^2$. This leads 
us to the conclusion that the flow equation \eq{flow} and 
an initial effective action $\Gamma_{k_0}$ 
given at a momentum scale 
$k_0$ may be used as a definition of the quantum theory. 

Now the question arises how to incorporate in this approach 
symmetries which are preserved on 
the classical level but are violated on quantum level. 
If the bare propagator $P^{-1}_\Phi$, the full propagator 
$(\Gamma_k^{(2)}[\Phi]+R^\Phi_k[P_\Phi])^{-1}$ and the 
regulator $R_k^\Phi$ are invariant or 
transform covariantly under 
a set of (global) symmetry transformations, the flow equation will 
preserve this symmetry. It is of course simple to introduce symmetry breaking 
via the regulator. However this makes it 
hard or even impossible to distinguish between effects 
due to this explicit symmetry breaking and quantum effects at intermediate 
steps and for finite $k$. Additionally symmetry breaking due entirely to the 
regulator should disappear in the limit where the cut-off is removed. Hence 
the symmetry breaking has to be introduced on the level of the initial 
effective action. Therefore a suitable approximation of 
$\Gamma_{k_0}$ should contain not only all terms with relevant 
couplings but as well the leading order in $1/k_0$ of
terms which break the symmetry of the classical action due to quantum
effects. Further terms are neglectable in $\Gamma_{k_0}$ since corrections due 
to them should 
remain of the order $1/k_0$ by 
means of universality even for lower scales $k$. 
Note that this is only valid for the {\em initial} 
value of the related couplings.

In the present case 
the regulator $R_k^\Phi$ and the bare propagator $P^{-1}_\Phi$ are 
invariant under global $U_A(1)$-variations if the effective action
$\Gamma_{k_0}$ at the starting point $k_0$ is $U_A(1)$-invariant. Thus no
$U_A(1)$-violating terms would arise during the flow. However we know from 
the $U_A(1)$-anomaly equation that 
$U_A(1)$ is broken on the quantum level. As explained above 
we have to introduce $U_A(1)$-violating terms to the effective action 
at the initial scale $k_0$ even though the 
couplings of these terms seem to be irrelevant at large scales.

$U_A(1)$-violating 
terms due to quantum effects have been derived in instanton calculations 
\cite{thooft2}. It is well known that the
semi-classical approximation used in instanton
calculations breaks down in the region where these effects become
important. In addition one has infrared problems due
to the integration over the zero modes (integration over the width of 
the instanton) (see \cite{shuryak,shifman} and references therein).  
We can deal with both problems in the 
framework of flow equations. The infrared cut-off ensures the infrared
finiteness and it is possible to go beyond the
semi-classical approximation by integrating the
flow equation to lower scales. The $U_A(1)$-violating terms calculated for 
the initial effective action $\Gamma_{k_0}$ act as the source for 
generating $U_A(1)$-violation in the flow equation via 
$(\Gamma_{k_0}^{(2)}+R_{k_0})^{-1}$. After integration of \eq{flow} 
they should be 
responsible for the anomalous
large mass of the $\eta'$-meson after integration. 

In the present
paper we concentrate on the derivation of the $U_A(1)$-violating
terms within the framework of the infrared regularised 
effective action. We show that the
fermionic sector contains zero modes generating 
$U_A(1)$-violating terms. The effective action is derived in an expansion in
orders $1/k_0$, valid at large cut-off scales $k_0$. 

In the second section we formulate the infrared regularised 
effective action within an
approach to flow equations developed in
\cite{reutwett} for pure Yang-Mills theory.
However we would like to emphasise that 
the results are independent of the approach chosen. 
We derive the effective action in the
zero instanton sector.  For later purpose the topological
$\theta$-term is included, where we choose the $\theta$-angle to
be space-time dependent (for a first discussion of the
$\theta$-parameter within the framework of flow equations see
\cite{reuter}). This is a natural extension, as the strong
$CP$-problem is deeply connected to the $U_A(1)$-problem. 
We show that the regularised effective action in the absence of 
zero modes (topological effects) interpolates 
between the classical action $(k\rightarrow \infty)$ and the 
full effective 
action ($k\rightarrow 0$). In the third section we calculate the 
$U_A(1)$-violating terms 
in the 1-instanton sector. In the fourth section we derive the 
effective
action at a large scale $k_0$ to leading order in $1/k_0$ by means
of universality. Multi-instanton contributions are of sub-leading order in
 $1/k_0$. In the last section we give a summary of the 
results and an outlook. In appendix \ref{noninvert} the existence of a
fermionic zero mode for the infrared regularised
Dirac operator is proven. The
appendices \ref{zeromode},\ref{local} contain some technical 
details.

\mysection{Derivation of the effective action}\label{sec2}
Throughout the paper we work in Euclidean space-time. In this 
section we 
only deal with topologically trivial configurations, 
since we want to derive the general properties of the effective
action in this theory. Hence the full propagators have only 
trivial zero modes.  
The classical action of a $SU(N)$-gauge theory coupled to 
fermions is given by  
\bm{c}\label{action}
S[A,\psi,\bar\psi]      =      S_A[A]+S_\theta[A]+S_\psi[A,
\psi,\bar\psi]
\eg 
where $S_A$ is the pure Yang-Mills action and 
$S_\psi$ is the fermionic action: 
\bm{ccc}\di   
S_A[A]  =  \frac{1}{2 g^2}\int\d^4 x\ \tr\, F^2(A), & &\di 
S_\psi[A,\psi,\bar\psi] =       \int \d^4 x\ \bar\psi
{\dr}(A)\psi.  
\eg 
$g$ is the coupling, 
the trace $\tr$ denotes a sum over Lorentz indices and the trace in the 
fundamental representation of the Lie algebra, 
$\tr\, t^a t^b =-\frac{1}{2}\delta^{ab}\delta_{\mu\mu}$. The
 field strength is given by 
\bm{ccccc} 
F_{\mu\nu}^a(A) =\partial_\mu A_\nu^a-\partial_\nu 
A_\mu^a+{f^a}_{bc}
A_\mu^b A_\nu^c, & &\di F=F^a t^a, & &\di [t^a,t^b]={f_c}^{ab} t^c. 
\eg 
The fermions $\psi=(\psi_{s,\xi}^A)$ 
are in the fundamental representation, where $A$ denotes the 
gauge group indices, $s=1,...,N_f$ the flavors and $\xi$ the spinor indices.  
The Dirac operator is given by 
\bm{ccccc}\label{dirac}
\dr_{\xi\xi'}^{AB}(A)=  (\gamma_\mu)_{\xi\xi'} D^{AB}_\mu, & &\di 
 D^{AB}_\mu(A)=  \partial_\mu\delta^{AB}+A^c_\mu (t^c)^{AB},
& &\di \{\gamma_\mu,
\gamma_\nu\}=-2\,\delta_{\mu\nu}.
\eg 
We have also introduced the topological $\theta$-term in \eq{action} 
\bg 
S_\theta[A] = \frac{i}{16\pi^2}\int\d^4 x\ \theta(x)
\tr\, F\tilde F(A), 
\eg 
where $\tilde F$ denotes the dual field strength. 
$S_\theta$ depends only on global properties of the gauge field for constant
$\theta$. In this case we may lose differentiability of $S_\theta$ 
with respect to the
gauge field. The effective action $\Gamma_k$ is defined by the
Legendre transformation of the infrared regularised 
Schwinger functional $W_k$. For
non-differentiable $W_k$ one has to use the general definition of the
Legendre transformation
\bg
\Gamma_k[\Phi]  = \inf_{J_\Phi}\left\{\int\d^4 x\, J_\Phi\Phi-
W_k[J_\Phi]\right\}.
\eg
However, it is difficult to use this definition for practical 
purpose. We circumvent this problem by allowing for an 
$x$-dependent $\theta$.
$\theta$ can be seen as a source of the index \cite{reuter}. 
              
 We derive the effective action $\Gamma_k$ for 
this theory within the path integral formalism. 
The effective actions for gauge theories and for fermionic 
systems in the present approach have been 
derived elsewhere (\cite{reutwett,bonetal,ellwanger,jungwett,bornwett}), so 
it is sufficient to outline this computation. We deal with the 
gauge degrees of freedom by  performing the Fadeev-Popov procedure. 
Given a gauge fixing condition $\CF^a[A,\bar A]=0$ we introduce 
\bm{ccc}\label{gauge}
 S_{g.f.}[A,\bar A]= \frac{1}{2\alpha}\int\d^4 x\, \tr\, \CF^2[A,\bar A], 
& &\di 
\CM^{ab}[A,\bar A](x,y)=D_\mu^{ac}(y)\frac{\delta}{\delta A_\mu^c(y)}
\CF^b[A,\bar A](x),  
\eg 
where $D_\mu^{ac}$ is the covariant derivative in the adjoint representation. 
In \eq{gauge} we have introduced the background field $\bar A$ 
(e.g.\ \cite{abbott}). Within the background field approach to gauge 
theories the generating functional of connected Green functions is 
given by 
\bm{rl}\label{def of W}
\exp W[J,\bar A,\eta,\bar\eta,\xi,\bar\xi] = &\di \!\! 
\frac{1}{\cal N}\int\CD a(c,\bar c)
\,\d\bar\chi\,\d\chi \exp\Bigl\{-S[a+\bar A,
\chi,\bar\chi]-S_{g.f.}[a,\bar A]\\\di 
&\di +\int_x\left(\bar\eta_{s,\xi}^A\chi_{s,\xi}^A+\bar
\chi_{s,\xi}^A \eta_{s,\xi}^A+  J_\mu^a a_\mu^a+
(\bar\xi^a c^a+\bar c^a\xi^a)\right)
\Bigr\},  
\eg 
where $\CN$ is an appropriate normalisation and the integral in the source 
terms is over space-time ($\int_x=\int \d^4 x$). Note also 
that the source $J$ only couples to the fluctuation field $a$. 
The gauge field measure $\CD a(c,\bar c)$ in \eq{def of W} 
depends on the chosen gauge and includes the ghost terms: 
\bm{c}
\CD a(c,\bar c) = \d a\,\d\bar c\,\d c\, \exp\left\{-\int_{x,y} \bar c^a(x) 
\CM^{ab}[a,\bar A](x,y) c^b(y)\right\}.  
\eg 
The interesting quantity is 
the generating functional of 1PI Green functions $\hat\Gamma$ 
which is the Legendre transformation of $W$. 
For the sake of convenience we introduce a shorthand notation for 
contractions, denoting them with a dot whenever the meaning is clear. (e.g.
\ $J\cdot a=J_\mu^a a_\mu^a,\ \bar\eta\cdot\psi=
\bar\eta_{s,\xi}^A\cdot \psi_{s,\xi}^A$). With this abreviation we get 
\bg
\hat\Gamma[A,\bar A,\psi,\bar\psi,\rho,\bar\rho]   =  -W[J,\bar
A,\eta,\bar\eta]+
\int_x\left(\bar \eta\cdot \psi+\bar\psi\cdot \eta+ J\cdot A+ 
(\bar\rho\cdot\xi+\bar\xi\cdot\rho)\right)
\eg
with
\bmo{ccccccccc}
\psi_{s,\xi}^A=\frac{\delta}{\delta\bar\eta_{s,\xi}^A} W, & &\di   
\bar\psi_{s,\xi}^A= -\frac{\delta}{\delta\eta_{s,\xi}^A} W, & &\di   A_\mu^a = 
\frac{\delta}{\delta J_\mu^a} W, & &\di 
\rho^a=\frac{\delta}{\delta\bar\xi^a} W, & &\di   
\bar\rho^a= -\frac{\delta}{\delta\xi^a} W. 
\ego 
In the following we concentrate on properties of 
the fermionic integration with the ghosts only being spectators. Thus 
for the sake 
of simplicity we will perform the calculation with vanishing ghost fields 
in the effective action, $\rho,\bar\rho=0$ ($\xi,\bar\xi=0)$. 
The further use of the Schwinger functional $W$ and the 
effective action $\hat\Gamma$ 
has to be understood in this sense. 
Using (\ref{def of W}), $\hat\Gamma$ is defined implicitly by the 
integro-differential equation 
\bm{rl}\label{funcimplicit}
\exp -\hat\Gamma[A,\bar A,\psi,\bar\psi] = &\di    
\!\!\frac{1}{\CN}\int\CD a\,\d\bar\chi\,\d\chi\exp\Biggl\{-S[a+\bar A,
\chi,\bar\chi]-S_{g.f.}[a,\bar A]\\\di 
&\di +\int_x\left(\frac{\delta\hat\Gamma}{\delta A}\cdot (a-A)
-\frac{\delta\hat\Gamma}{\delta
  \psi}\cdot(\chi-\psi)+(\bar\chi-\bar\psi)\cdot 
\frac{\delta\hat\Gamma}{\delta\bar\psi}\right)\Biggr\}, 
\eg
where we have used 
\bmo{ccccc}
\eta_{s,\xi}^A=\frac{\delta}{\delta\bar\psi_{s,\xi}^A}\hat\Gamma, & &\di   
\bar\eta_{s,\xi}^A=  -\frac{\delta}{\delta \psi_{s,\xi}^A}\hat\Gamma, 
& &\di   J_\mu^a=\frac{\delta}{\delta A_\mu^a}\hat\Gamma. 
\ego 
However, these expressions are only formal notations and do not make sense 
without a suitable regularisation. The regularisation is introduced in
the following way: We use an 
explicit infrared regularisation dependent on the cut-off scale $k$ 
(see \eq{cut-off1},\eq{limits}). 
The flow equation \eq{flow} describes the
$k$-dependence of $\Gamma_k$. 
The ultraviolet regularisation, dependent on the cut-off parameter
$\Lambda$, is only used 
implicitly. We allow for the class of ultraviolet regularisations 
preserving the gauge symmetry. These regularisations violate the axial
symmetry due to a general no-go theorem (see e.g.\ \cite{current}). 
Within the framework of flow equations this is
equivalent to a fixing of the (axial) anomaly by choosing appropriate 
ultraviolet boundary conditions \cite{bondat}.

The cut-off term for the gauge-field is introduced as 
follows:  
\bm{ccc}\label{gaugecut} 
 \Delta_k S_A[a,\bar A] =  -\frac{1}{2}\int_x a\cdot  
R_{k}^{A}[D_T(\bar A)]\cdot a, & \&\di 
R_{k\mu\nu}^{A,ab}[D_T] =  \left(D_T
\frac{1}{e^{D_T/k^2}-1}\right)^{ab}_{\mu\nu}.  
\eg 
The operator $D_T$ is defined by  
\bm{c}
D_{T,\mu\nu}^{ab} =    -D^{ac}_\rho D^{cb}_\rho\delta_{\mu\nu}-
2g {f_c}^{ab}F^c_{\mu\nu}  
\eg 
where $D^{ab}_\rho$ is the covariant derivative in the adjoint representation. 
For the details of the background field approach to flow equations we
 refer the reader to \cite{reutwett}. 
The infrared cut-off for the fermions is defined in a similar way. 
As discussed in the introduction
 we use a fermionic cut-off term having the same 
Dirac structure as the inverse of the bare fermionic
propagator. Hence it is proportional to $\dr(\bar A)$. We would like to 
emphasise that the results 
obtained here easily extend to more general cut-off terms. 
However in the general case 
the technical details are more involved. Thus a convenient choice is  
\bm{cc}\label{fermcut}
\Delta_k S_\psi[\psi,\bar\psi,\bar A] = \int_x 
\bar\psi\cdot R_k^\psi[\dr(\bar A)]\cdot\psi, &\di    
R_{k,\xi\xi'}^{\psi,AB}[\dr(\bar A)] =   \Biggl(\dr(\bar A)
\frac{1}{\Bigl(e^{\Dr^2(\bar A)/k^2}-1\Bigr)^{1/2}}\Biggr)_{\xi\xi'}^{AB}. 
\eg 
The cut-off terms (\ref{gaugecut},\ref{fermcut}) have the limits
presented in (\ref{limits}). We only refer implicitly to the cut-off term 
$R_k^c$ of the ghosts by keeping 
the gauge field measure $k$-dependent ($\CD a_k$). 
Note that although the ghosts are spectators in 
the present derivation $R_k^c$ leads to ghost contributions in the flow 
equation \eq{flow} (e.g.\ \cite{ellwanger}). 

The $\Lambda$-dependence of the effective action is not specified,
because we are working in the limit $\Lambda\rightarrow \infty$. In 
appendix~\ref{noninvert} we discuss the properties of
the ultraviolet cut-off in more detail. As a result 
the functional measures of the gauge field and the fermions are 
$\Lambda$-dependent. For the sake of brevity we will drop any reference to 
the ultraviolet cut-off. We get for the regularised Schwinger 
functional $W_k$   
\bm{rl}\label{def of W_k}
\exp W_k[J,\bar A,\eta,\bar\eta]= 
\frac{1}{\CN_k}\int\CD a_k\,\d\bar\chi\,\d\chi
\exp\Bigl\{-S_k[a,\bar A,
\chi,\bar\chi]+\int_x\left(\bar\eta\cdot\chi+\bar
\chi\cdot\eta+J\cdot a\right)\Bigr\}, 
\eg 
where the action $S_k$ and the normalisation $\CN_k$ are given by 
\bmo{ccccc}
S_k=   S+S_{g.f.}+\Delta_k S, & &\di \Delta_k S=\Delta_k S_A+
\Delta_k S_\psi, &  &\di 
{{\cal N}_k}  = \exp\left\{W_k[0,\bar A,
0,0]\right\}   .
\ego 
The limit $k\rightarrow \infty$ of the Legendre transformation 
of $W_k$ is divergent since it is 
proportional to $\Delta_k S$. However it is 
possible to define the effective action $\Gamma_k$ 
in a way that 
guarantees a finite $k\rightarrow \infty$ limit by  
\bm{c}\label{G_klegendre}
\Gamma_k[A,\bar A,\psi,\bar\psi] =   
-W_k[J,\bar A,\bar\eta,\eta]+\int_x\left(\bar \eta\cdot\psi+
\bar\psi\cdot\eta+J\cdot A\right)-\Delta_k S[A,\bar A,
\psi,\bar\psi].
\eg 
Now the regularised analogue of \eq{funcimplicit} follows immediately from 
\eq{def of W_k} and \eq{G_klegendre} as 
\bm{rl}\label{G_k}
\exp\left\{-\Gamma_k[A,\bar A,\bar\psi,\psi]\right\}
= &\di 
\!\!  \frac{1}{{\cal N}_k}\int\CD a_k
\,\d\bar\chi\,\d\chi
\exp\Bigl\{-S_k[a,\bar A,\chi,\bar\chi]+\Delta_k S[A,\bar A,\psi,
\bar\psi]\\\di  
    &\di  +\int_x\left(\bar\eta\cdot(\chi-\psi)+(\bar
\chi-\bar\psi)\cdot \eta  +J\cdot(a-A) \right)\Bigr\}. 
\eg
The sources $(\eta,\bar\eta,J)$ in \eq{G_k} have to be understood as 
functions of $\Gamma_k$ and follow from \eq{G_klegendre} as  
\bm{ccccc}\label{sources}
\eta_{s\xi}^A=\frac{\delta}{\delta\bar
\psi_{s,\xi}^A}(\Gamma_k+\Delta_k S_\psi), & &\di   
\bar\eta_{s,\xi}^A=  -\frac{\delta}{\delta 
\psi_{s,\xi}^A}(\Gamma_k+\Delta_k S_\psi), & &\di   J_\mu^a
=\frac{\delta}{\delta A_\mu^a}(\Gamma_k+\Delta_k S_A). 
\eg 
Hence $\Gamma_k$ is implicitly defined by \eq{G_k}. 
The effective action can be written in a more convenient form by 
using the following shift of variables: 
\bm{ccccc}\label{shift}
a \rightarrow a-A, & &\di    \bar\chi\rightarrow
\bar\chi-\bar\psi, & &\di   \chi\rightarrow\chi-\psi.
\eg
Applying \eq{shift} to \eq{G_k} we get 
\bm{c}\label{G_k1}
\exp\left\{-\Gamma_k[A,\bar A,\psi,\bar\psi]\right\}
=\frac{1}{\CN_k}\int\CD a_k\,\d\bar\chi
\,\d\chi\exp\Bigl\{-S[a+\bar
A+A,\bar\chi+\bar\psi,\chi+\psi]\\\di 
  -S_{g.f.}[a+A,\bar A]-\Delta_k S[a,\bar A,\chi,\bar\chi]
+\int\left(\bar\eta\cdot\chi+\bar\chi\cdot 
\eta+J\cdot a\right)\Bigr\}.
\eg 
In the limit 
$k\rightarrow \infty$ the path integral is dominated by the 
$\Delta_k S$ terms and to leading order we can neglect the dependence 
of the other terms on the fields  
$a,\chi,\bar\chi$ (e.g.\ $S[a+A+\bar A,\chi+\bar\psi,\chi+\psi]
\rightarrow 
S[A+\bar A,\psi,\bar\psi]$). What 
is left is a Gaussian integral which is canceled by the 
normalisation $\CN_k$. Thus 
$\Gamma_k$ approaches the classical action for $k\rightarrow\infty$ 
(including the gauge fixing). In the limit
$k\rightarrow 0$ the 
$k$-dependence of $\Gamma_k$
vanishes and $\Gamma_k$ approaches the full effective action:  
\bm{c}\label{0infty}
S+S_{g.f.}\stackrel{k\to\infty}{\longleftarrow}
\Gamma_k \stackrel{k\to 0}{\longrightarrow} \Gamma.
\eg 
Hence we have derived the regularised effective action 
for a gauge theory coupled to fermions in the gauge field sector with trivial
topology. $\Gamma_k$ interpolates between the classical action
for $k\rightarrow\infty$ and the full effective action for $k= 0$.

\mysection{Instanton-induced terms}
The result \eq{0infty} of the last section remains valid in gauge field 
sectors with non-trivial topology. However we have to take into account also 
$U_A(1)$-violating terms which may be suppressed with powers of $1/k$. It can 
be shown that the Dirac operator has a non-trivial zero mode in the 
1 instanton sector even after the infrared regularisation 
(see appendix \ref{noninvert}). This serves as the source for 
$U_A(1)$-violating terms. For the calculation of these terms 
we consider gauge field configurations 
with instanton number $\pm 1$ and carefully examine 
the zero mode dependence of the path integral. 

For the non-zero modes the derivation of the last section
holds without modification. Due to the infrared
cut-off $\Delta_k S_A$ there are no gauge field zero modes. The scale
invariance of the action is broken and the minimum of the action is at
vanishing instanton width. 
This can be seen as follows: Let $a$ be a configuration with
instanton number $1$. Additionally we choose 
$A,\bar A$ to be in the trivial sector. It follows that 
\bm{ccc}
S_A [a+A+\bar A]\geq   \frac{8\pi^2}{g^2}, & &\di  
\Delta_k S_A[a,\bar A]\geq 0.
\eg
In addition, $\Delta_k S_A$ is vanishing only for gauge fields with
vanishing norm (see appendix~\ref{local}). Thus the gauge field
sector has no infrared problems even if topologically non-trivial gauge
field configurations are considered. In the limit 
$k\rightarrow\infty$ the gauge field integration becomes trivial. This remains
valid for $A,\ \bar A$ with 
arbitrary instanton number. To simplify the following
calculations we assume $A, \bar A$ to be in the trivial sector. First
 we have a closer look at the fermionic part
of the action. We shall argue by using the limit $\Gamma_k 
\stackrel{k\to\infty}{\longrightarrow} S+S_{g.f.}$ that
 only the source terms couple to the fermionic zero mode. Therefore 
the zero mode integration can be done explicitly. 
After shifting the fields as in 
(\ref{shift}) the part of the exponent in (\ref{G_k1}) 
depending on fermionic variables
reads for instanton configurations $a_I$
\bm{rl}\label{fermpart}
&\di\!\! -S_\psi[a_I+A+\bar A,\chi'+\psi',\bar\chi'+\bar\psi']-
\Delta_k
S_\psi[\chi'+\psi',\bar\chi'+\bar\psi',\bar A]\\\di 
&\di \!\! +\int_x\left(
\bar\eta\cdot\chi
+\bar\chi\cdot\eta\right)+\Delta_k S_\psi[\psi,\bar\psi,\bar A],
\eg 
where the primed fermionic fields are the non-zero modes of the infrared
regularised Dirac operator $\dr(a_I+A+\bar A)+\r_k[\dr(\bar A)]$ and 
the zero modes are denoted by $\chi_0,\psi_0$. In the limit 
$k\rightarrow\infty$ we get for the sources 
\bm{ccc}\label{classlim}
\eta  \rightarrow \frac{\delta }{\delta \bar\psi}(S_\psi+\Delta_k S_\psi), 
& &\di \bar\eta 
\rightarrow -\frac{\delta}{\delta \psi}( S_\psi+\Delta_k S_\psi). 
\eg 
Using \eq{classlim} we deduce from (\ref{fermpart})
\bm{rl}
 &\di 
\!\! -S_\psi[a_I+A+\bar A,\psi',\bar\psi']
-S_\psi[a_I+A+\bar A,\chi',\bar\chi']-\Delta_k
S_\psi[\chi',\bar\chi',\bar A]\\\di 
&\di \!\! +\int_x\left(\bar\eta\cdot\chi_0
+\bar\chi_0\cdot\eta\right)+\Delta_k S_\psi[\psi,\bar\psi,\bar A]-
\Delta_k S_\psi[\psi',\bar\psi',\bar A]. 
\eg 
The terms linear in $\bar\chi_0,\ \chi_0$ remain. There is no
counterterm in the action, since $S_\psi+\Delta_k S_\psi$ does not
depend on the zero mode. The cross terms $\Delta_k
S_\psi[\bar\chi',\psi',\bar A],
\ \Delta_k S_\psi[\bar\psi',\chi',\bar A]$ are
canceled by similar ones derived from the source terms. 
We have dropped the
cross terms $S_\psi[a_I+A+\bar A,\bar\chi',\psi'],\ S_\psi[a_I+A+\bar
A,\bar\psi',\chi']$, since they are suppressed with 
$1/k$ in the limit $k\rightarrow\infty$. We have, with $\dr\psi_0\ =\ -
\r_k\psi_0$, 
\bm{c}
S_\psi[a_I+A+\bar A,\psi',\bar\psi']-\Delta_k
S_\psi[\psi,\bar\psi,\bar A]+
\Delta_k S_\psi[\psi',\bar\psi',\bar A]=  
S_\psi[a_I+A+\bar A,\psi,\bar\psi].
\eg
The final result for the fermionic part of the exponent in \eq{G_k1} is 
\bm{c}\label{finalres}
 -S_\psi[a_I+A+\bar A,\psi,\bar\psi]
-S_\psi[a_I+A+\bar A,\chi',\bar\chi']-\Delta_k
S_\psi[\chi',\bar\chi',\bar A]
 +\!\int_x\left(\bar\eta\cdot\chi_0+\bar\chi_0\cdot 
\eta\right)
\eg 
The last term in \eq{finalres} depends on the 
instanton $a_I$ via the zero mode. However only 
instantons $a_I$ with width $\rho\sim1/k$ contribute. The 
infrared regularisation of the gauge field supresses instantons 
with width $\rho\gg 1/k$ (see appendix \ref{local}). 
We split the gauge 
field measure into the measure of collective coordinates of the
instanton and the measure of fluctuations about the instanton. 
Let $\d a_{I,k}$ 
be the ($k$-dependent) measure of the collective 
coordinates of the $SU(N)$-instanton \cite{thooft2,bernard}. 
The zero mode contribution 
factorises in the limit
$k\rightarrow\infty$. 
Hence taking into account the trivial sector and the $\pm 1$ instanton 
sectors the effective action is given by 
\bm{rl}\label{zero:01}
\exp\left\{-\Gamma_k[A,\bar A,\psi,\bar\psi]\right\}
 =&\di \!\!\exp\left\{-S[A+\bar
  A,\psi,\bar\psi]+S_{g.f.}[A,\bar A]\right\}\Bigl(1+
 \Bigl[\int\d\mu_1(\theta)\,\d \bar\chi_0\,\d \chi_0\\\di 
&\di \times\exp\int\left(\bar\eta\cdot\chi_0 
+\bar\chi_0\cdot\eta\right)+\mbox{h.c.}\Bigr]\Bigr)+O(1/k)
\eg 
with 
\bm{ccc}\label{nkprime} 
\d\mu_1(\theta)=   \d a_{I,k} \frac{{\CN_k}'[a_I]}
{\CN_k}, &  &\di  
{\CN_k}' [a_I]  =  \int\CD a_k'
\,\d\bar\chi'
\,\d\chi' \exp\left\{-S_k[a+a_I,0,\chi',\bar\chi']\right\}.
\eg 
We have dropped the dependence of the zero
mode contribution on $A$ and $\bar A$, since it is 
only next to leading order in $1/k$ (see appendix 
\ref{zeromode}). We also have used that the 
contribution from the sector with 
instanton number $-1$ is the hermitean conjugate of the sector with 
instanton number $1$. Thus we concentrate on the 
sector with instanton number $+1$ and compute 
\bm{c}\label{term}
\int\prod_{s=1}^{N_f}\d \bar b_0^s\,\d a_0^s
\exp\int_x\left(\bar\eta\cdot\chi_0+\bar\chi_0\cdot
\eta\right)= \prod_{s=1}^{N_f}\left(\int_x
\bar\eta_s\cdot\phi_{0}\right)\,\left(\int_x \phi_{0}^+
\cdot\eta_s\right)
\eg 
with 
\bm{cccc}
(\chi_0)_{s,\xi}^A =   a_0^s\,\phi^A_{0,\xi},&\di   
(\bar\chi_0)_{s,\xi}^A =   (\phi_0^+)_\xi^A\,\bar b_0^s, &\di   
\int_x \phi^+_0\cdot\phi_0=1, &\di \d\bar\chi_0\,\d\chi_0=
\prod_{s=1}^{N_f}\d \bar b_0^s\,\d a_0^s.  
\eg
Higher powers of $(\int\bar\eta\cdot \chi_0)(\int\bar\chi_0\cdot\eta)$ 
vanish because of the properties of Grassmann variables. These 
calculations result in an effective
action $\Gamma_k$, which is given in terms of 
an integro-differential equation even 
in the limit $k\rightarrow\infty$ as opposed to \eq{0infty}. 
\bm{c}\label{implicit}
\Gamma_k[A,\bar A,\psi,\bar\psi] 
\stackrel{k\rightarrow\infty}{\longrightarrow}
S[A+\bar
A,\psi,\bar\psi]+S_{g.f.}+P_k[A,\bar A,\psi,\bar\psi]
+O(1/k), 
\eg 
where the $U_A(1)$ violating term $P_k$ is 
\bm{c}\label{P_k}
P_k[A,\bar A,\psi,\bar\psi] = \int \d\mu_1(\theta)
\prod_{s=1}^{N_f}\left(\int_x\bar\eta_s\cdot 
\phi_0\right)\left(\int_x\phi^+_0\cdot\eta_s
\right)  +\mbox{h.c.}
\eg 
The sources $\eta,\bar\eta$ are given 
in \eq{sources} as functional derivatives of $\Gamma_k$ with respect to 
$\psi,\bar\psi$. Terms which are suppressed with 
powers of $1/k$ but do not violate the $U_A(1)$ are contained in 
$O(1/k)$. Although we will see in the
next section that 
$P_k$ is also suppressed with powers of $1/k$, it is not
possible to neglect it since it introduces $U_A(1)$-violation.

\mysection{Effective action in the large scale limit} 
Equation (\ref{implicit}) is a
functional differential equation for
$\Gamma_k$. In the limit $k\rightarrow\infty$ we are able to solve 
this equation. First we note that for $k\to\infty$ the $U_A(1)$-violating term 
takes a local form. The explicit calculation
is given in appendix~\ref{zeromode}. As mentioned before, we
do not have gauge field zero modes. 
For quantitative purposes one should work within the valley 
method (see \cite{shifman} and references therein). 
However for our purpose it is enough to estimate of the value of 
the coupling of the $U_a(1)$-violating term. 
As we will see later on only instantons with width 
$\rho\sim 1/k\to 0$ contribute. The qualitative behaviour of the  
corresponding fermionic zero modes does not change in the presence of the
 cut-off term. The zero modes have width $\rho\to 0$ and are peaked about 
the centre of the instanton. It follows (see appendix
\ref{zeromode}, (\ref{split},\ref{groupint2},\ref{finresult}))
\bm{c}\label{quote}
P_k[\psi,\bar\psi] = \int_z\Delta[k,\theta]\det_{s,t}\bar\eta_s(z)
\frac{1-\gamma_5}{2}\eta_t(z)+\mbox{h.c.}+O(\Delta[k,\theta]/k)
\eg 
with 
\bm{ccc}  
\Delta[k,\theta]  =   (2^5\pi^2\rho^4)^{N_f} \int
\d\bar\mu_1(\theta)\,a[N,N_f]\ \sim  k^{-5N_f+4}. 
\eg 
$P_k$ does not depend on $A,\ \bar A$ to leading order 
(see appendix~\ref{zeromode}), so the $U_A(1)$-violation is purely fermionic 
to leading order. Now we concentrate on the measure $\d\bar\mu_1(\theta)$. 
The fluctuation fields $a',\chi',\bar\chi'$ decouple approximately
 from the instanton for large scales $k$ (see discussion about the use 
of the valley method). This can be used to effectively remove the 
gauge fixing term for $a_I$. 
We have in the limit $k\to\infty$ (see (\ref{nkprime}))
\bm{c}\label{ratio}
\frac{{\CN_k}'[a_I]}{\CN_k} 
 \sim \frac{1}{\CN_k}\int\CD a_k'
\,\d\bar\chi'
\,\d\chi' e^{-S_k[a',0,\chi',\bar\chi']}
\exp\left\{-\Delta_k S_A[a_I,0]
-S_A[a_I]-S_\theta[a_I]\right\},
\eg 
where the measures do not include the zero (or quasi-zero) 
modes related to $a_I$. Note for explicit calculations that the gauge 
field integration also includes the ghost contribution with a 
suitable regulator $R_k^c$ for the ghosts. The integrals in \eq{ratio} 
become Gaussian for $k\to\infty$. Taking into account the normalisation 
$\CN_k$ they lead to a factor $\rho^{n_0^f-n_0^g}$, where $n_0^f,\ n_0^g$ 
are the number of fermionic 
(f) zero modes and corresponding gauge field (g) modes and $\rho$ is the 
width of the instanton. The factor $\exp\{-\Delta_k S_A[a_I,0]\}$ provides an 
exponential suppression of the zero mode contribution for $\rho\gg 1/k$ 
due to the infrared regularisation of the gauge field \eq{explicit}. 
This ensures the infrared 
finiteness of the $\rho$ integration. Therefore we can assume 
$\rho$ to be of order $1/k$ or smaller. The term  
\bm{c}
\exp\left\{-S_A[a_I]\right\}= \exp\left\{-
\frac{8\pi^2}{g^2}\right\}
\eg
is well known from instanton calculations (see \cite{shifman} and
references therein). The exponent $S_\theta$ of the
remaining 
factor is related to the instanton number 
$\frac{1}{16\pi^2}\int \tr\,F\tilde F=1$. In the limit $\rho\to 0$ 
the density $\tr\, F\tilde F[a_I(x)]$ serves as a $\delta$-function which is 
peaked at the centre $z$ of the instanton. Hence we get for 
$\rho\sim 1/k\to 0$ 
\bm{c} 
\exp\left\{-\frac{i}{16\pi^2}\int_x \theta (x)\tr\, F\tilde F[a_I(x)]
\right\}\rightarrow \exp\left\{-i\theta(z)
\frac{1}{16\pi^2}\int_x \tr F\tilde F[a_I]\right\} 
=\exp\{-i\theta(z)\}.
\eg 
Thus the $\theta$-term leads to the following modification
of the $\det_{s,t}$-term:
\bm{c}
\int_z\Delta[k,\theta]\det_{s,t}\bar\eta_s\frac{1-\gamma_5}{2}\eta_t 
 = \int_z
\Delta[k,0]e^{-i\theta(z)}\det_{s,t}\bar\eta_s
\frac{1-\gamma_5}{2}\eta_t(1+O(1/k)). 
\eg
Anti-instantons  have instanton number $-1$, 
so in this case one picks up a factor $\exp\{i\theta(z)\}$. 
Using these results in (\ref{implicit}) we end up with 
\bm{c}\label{implicit2}
\Gamma_k[A,\bar A,\psi,\bar\psi] = S[A+\bar
A,\psi,\bar\psi]+S_{g.f.}[A,\bar A]
+P_k[\psi,\bar\psi]+O(1/k)
\eg
with 
\bm{c}\label{P_k1}
P_k[\psi,\bar\psi]  = \int_z\Delta[k,\theta] \det_{s,t}\left[
-\frac{\delta}{\delta \psi_s}(
\Gamma_k+\Delta_k S_\psi)
\frac{1-\gamma_5}{2}\frac{\delta}{\delta\bar\psi_t}(
\Gamma_k+\Delta_k S_\psi)\right]+\mbox{h.c.} 
\eg
In \eq{P_k1} we have used the explicit dependence of $\eta,\bar\eta$ on 
$\Gamma_k$ as given in \eq{sources}. 
The term $O(1/k)$ includes sub-leading orders
 of $U_A(1)$-conserving contributions and
$U_A(1)$-violating contributions. 
The factor $\Delta[k,\theta]$ provides a suppression of $P_k$
proportional to $k^{-5N_f+4}$. 
          
 The properties of (\ref{implicit2}) lead to an effective
action $\Gamma_k$, which is well-defined in the limit $k\rightarrow
\infty$. In addition an explicit expression for $\Gamma_k$ can be
derived. Note that we have used in the derivation 
that \eq{0infty} is also valid in the instanton sectors. Hence 
proving the existence of a well-defined limit of $\Gamma_k$ 
serves as a self-consistency check. The only source for a
diverging contribution is $P_k$, which is purely fermionic. In the
limit $k\rightarrow \infty$ we have (see (\ref{fermcut}))
\bm{ccc} \label{limes}
\frac{\delta}{\delta\bar\psi}\Delta_k S_\psi  \rightarrow
k\frac{\p}{|\p|}\psi, &\di &\di 
-\frac{\delta}{\delta \psi}\Delta_k S_\psi  \rightarrow k
\bar\psi \frac{\p}{|\p|}.
\eg 
Combined with $\Delta[k,\theta]$ these terms are still suppressed with
powers of $1/k$ $(N_f>1)$. In addition, (\ref{implicit2}) is 
inconsistent for 
\bm{c}\label{glim}
\frac{\delta}{\delta \psi}\Gamma_k,\
\frac{\delta}{\delta\bar\psi}\Gamma_k\sim k^n, \ \ \ \ 
\ n\neq 0.  
\eg
This follows by using 
$\Delta[k,\theta]\sim k^{-5 N_f+4}$ and (\ref{limes}). The only consistent 
choice in \eq{glim} is $n=0$ which also 
ensures the finiteness of $\Gamma_k$ in the limit $k
\rightarrow\infty$. Thus we drop contributions  in $P_k$ 
dependent on $\frac{\delta}{\delta \psi}\Gamma_k,\
\frac{\delta}{\delta\bar\psi}\Gamma_k$ since they are of 
sub-leading
order. With \eq{limits} and \eq{fermcut} we get 
\bm{c}
\bar\psi_s\r_k \frac{1\pm\gamma_5}{2}\r_k \psi_t  =
\bar\psi_s{\r_k}^2
\frac{1\mp\gamma_5}{2}\psi_t
  \stackrel{k\rightarrow \infty}{\longrightarrow}  k^2 
\bar\psi_s
\frac{1\mp\gamma_5}{2}\psi_t
\eg
and the final result for the effective action for large scales $k$ is  
\bm{rl}\label{final}
\Gamma_k[A,\bar A,\psi,\bar\psi] =   &\di \!\! S[A+\bar
A,\psi,\bar\psi]+S_{g.f.}[A,\bar A]+\int_z\Delta[k,\theta] k^{2N_f}\det_{s,t}
\bar\psi_s
\frac{1+\gamma_5}{2}\psi_t\\\di  
  &\di    +\int_z\Delta^*[k,\theta] k^{2N_f}\det_{s,t}\bar\psi_s
\frac{1-\gamma_5}{2}\psi_t+O(1/k,\Delta[k,\theta]k^{2N_f-1}).
\eg
with $\Delta[k,\theta]k^{2N_f}\sim k^{-3 N_f+4}$ and $\Delta^*[k,\theta]$ 
is the 
hermitean conjugate of $\Delta[k,\theta]$. 
The term $O(1/k,\Delta[k,\theta]k^{2N_f-1})$ 
includes sub-leading orders of $U_A(1)$-conserving contributions 
(first argument) and $U_A(1)$-violating contributions (second argument). 

In (\ref{final}) the contributions of the trivial sector and the sector
with instanton number $\pm 1$ are included. However, contributions
$P_k(n)$ of sectors with instanton number $n,\  |n|\geq 2$ are only 
sub-leading terms in $1/k$. 
Due to the cut-off term for the gauge field (\ref{gaugecut}) 
the contributions exhibit a natural size $\approx 1/k\to 0$. 
This locality is sufficient to allow qualitatively 
the same arguments as in the derivation of the
$U_A(1)$-violating terms in the one instanton sector and end up with
powers of the flavor determinant 
\bm{c}
P_k(n) \sim k^{-3n\cdot N_f+4}\int_z 
\left(\det_{s,t}\bar\psi^{A_s}_s\frac{1-\gamma_5}{2}\psi^{B_t}_t\right)^n 
\CT_n^{A_1 B_1\cdots A_{n N_f}B_{n N_f}}, 
\eg
where $\CT_n$ denotes the color structure. These are sub-leading terms.

\mysection{Discussion}
We have calculated the fermionic $U_A(1)$-violating terms
 contributing to the effective action $\Gamma_k$ in the presence of 
instantons to leading order in $1/k$. Due to the
infrared cut-off term of the gauge field there are no problems with
infrared divergences, so there are no gauge field zero modes. 
Although we have introduced infrared cut-off terms
for the fermionic fields, there is still a fermionic zero mode for
instanton configurations.  For this reason the integration of the
fermionic zero mode sector factorises and we end up with the well
known 't~Hooft determinant as the first order correction in $1/k$. The
coupling $\Delta[k,\theta]$ of the 't~Hooft determinant is infrared
finite due to the gauge field regularisation. Further
corrections are of sub-leading order in $1/k$. In addition they show
the same flavor structure as the 't Hooft determinant. Inclusion of fermionic 
mass terms is straightforward. The effective action (\ref{final}) 
with suitable wave function 
renormalisations, an explicit ghost sector (see
\cite{reutwett,ellwanger}) and additional $U_A(1)$-conserving terms 
may serve as an appropriate input to the exact flow
equation in order to study the $U(1)$-problem. 

It is possible to calculate $\mu[k,\theta]$ numerically based on the
results of appendix \ref{zeromode}. The result does not allow
quantitative statements, but provides a validity bound of
the $1/k$-expansion.  If the expansion is still valid at
about $k\sim 700$ MeV, it would be possible to take the value of
$\mu[k,\theta]$ at $k\sim 700$ MeV as an
input for a phenomenological quark-meson model (see \cite{jungwett}). 
We briefly discuss the approximations used in the numerical
calculation and comment on the result. The renormalisation scheme
proposed by the framework of flow
equations (with appropriate boundary conditions) can be related to the 
$\overline{\mbox{MS}}$-scheme \cite{ellwanger2}. Therefore 
one works in a first approximation with the well known results at
one-loop level of instanton calculations and introduces as the only new
ingredient the infrared cut-off in the gauge field sector. 
Clearly these approximations 
allow only a rough estimate of the value of $\mu[k,\theta]$. Moreover 
one can determine the validity-bound of the $1/k$-expansion in
the case of QCD. For scales 
$k\sim 1-1.3$ GeV the $1/k$ approximation breaks down, and one has to
use the flow
equation to extrapolate to lower energies. It is
known from instanton calculations, that in this region
corrections which are proportional to the 
gluonic condensate $\langle 1/g^2\cdot F^2\rangle$ become important (see 
\cite{shuryak,shifman}). Since the flow should be 
smooth (as opposed to the case of correlation functions connected with 
phase transitions), one expects 
fewer problems with the numerical integration of the flow equation 
for the relevant correlation functions (e.g.\ $\Delta[k,\theta]$, which
is connected to the $\eta'$-mass). On the other hand the
value of $\Delta[k,\theta]$ in the physical region should be 
dominated by the contributions collected during the flow, otherwise the 
value would depend on the initial scale $k_0$, which has no physical meaning.
Therefore the calculation of correlation functions connected with the
instanton-induced terms is interesting for two reasons: 
It is a good check for the computational power of flow equations and 
it would be a great success to derive the $\eta'$-mass
quantitatively from first principles. 

We have also discussed the leading order corrections due to the 
$\theta$-term. It leads to an additional phase factor in
the 't Hooft determinant breaking $CP$-invariance in the presence of massive 
fermions. In the 
case of massless fermions the $\theta$-angle can be absorbed in a
redefinition of the fields. For massive fermions
 one can calculate the flow of $\theta$. 
To solve the strong $CP$-problem, one has to 
calculate $\theta$ in the full quantum theory. For pure QCD the 
flow of $\theta$ has been studied in \cite{reuter} using a rough 
approximation. Also fermionic contributions to the flow in a 
one-flavor model have been discussed. 
These calculations indicate a non-trivial scale
dependence of the $\theta$-parameter. In addition it is shown that 
the infrared limit $k\rightarrow 0$ has to be studied to give an final
answer to this problem. For this issue one needs both 
a truncation of the flow equation leading to a consistent flow 
in the regime of interest and an initial effective action $\Gamma_{k_0}$ 
including all important terms. Certainly the $U_A(1)$-violating terms 
derived in the present paper are necessary initial inputs. 

\subsection*{Acknowledgments}
I would like to thank C.~Wetterich for many helpful discussions 
and A.~Wipf for discussions to appendix \ref{noninvert}.

\begin{appendix}
\mysection{Zero modes of the regularised Dirac operator}
\label{noninvert}
In this appendix we discuss the behaviour of the regularised fermionic 
functional integral. We are interested in 
\bm{c}\label{unreg}
Z_{\psi,k}[a,\bar A,\bar\eta,\eta]=\frac{1}{\CN_k}
\int\d\bar\chi\,\d\chi 
\exp\int_x\left\{-\bar\chi(\dr(a+\bar A)+\r_k[\dr(\bar A)])\chi
+\bar\eta\cdot\chi+\bar\chi\cdot \eta\right\},
\eg
with the normalisation given by $\CN_k=Z_{\psi,k}[0,0,0,0]$. 
The Dirac operator $\dr[a+\bar A]$ has one zero
eigenvalue (non-degenerate) for a configuration $a+\bar A$ with
instanton number $\pm 1$. First we discuss the ultraviolet regularisation of 
\eq{unreg}. For this purpose we concentrate on 
\bm{c}\label{zeta}
Z_{\psi,k}[a,\bar A,0,0] =  
\left(  
\frac{\det_\Lambda\left(\dr(a+\bar A)+\r_k[\dr(\bar A)]\right)}{
\det_\Lambda\left(\dr(0)+\r_k[\dr(0)]\right)}\right)^{N_f}.
\eg
The subscript $\Lambda$ is related to the fact that an 
ultraviolet regularisation of the determinants is needed. An
appropriate regularisation in (\ref{zeta})
would be the $\zeta$-function regularisation. More generally, high 
momenta should be suppressed in a gauge-invariant way. 
These conditions are fulfilled by the 
regularisations 
$g_\Lambda[\dr(a+\bar A)+\r_k[\dr(\bar A)]]$ of the Dirac operator with the 
properties 
\bm{ccccc}\label{uvconditions}
g_\Lambda[0]   =  0,& &\di g_\Lambda[x]  
\stackrel{x^2\gg\Lambda^2}{\longrightarrow} 0, &  
&\di \left\{g_\Lambda[x],\gamma_5\right\} =0\ \  \mbox{if} \ \ 
\{x,\gamma_5\} =  0.
\eg 
An explicit $g_\Lambda$ fulfilling (\ref{uvconditions}) is 
\bm{c}
g_\Lambda[x] =x e^{-x^2/\Lambda^2}.
\eg 
$g_\Lambda$ does not influence the infrared behaviour of the Dirac
operator. In particular it vanishes for zero modes. 
Hence we use $\dr(a+\bar A)+\r_k[\dr(\bar A)]$ for the
discussion of the zero mode. $\dr(a+\bar A)+\r_k[\dr(\bar A)]$ 
is not invertible on the 
one-dimensional 
subspace of the zero mode $\chi_0$ of $\dr(a+\bar A)$, i.e.\ acting with 
the inverse of $\dr(a+\bar A)+\r_k[\dr(\bar A)]$ 
on $\chi_0$ does not lead to a square-integrable function. 
This indicates the existence of a zero mode. 
To prove the existence of a zero mode for the infrared regularised 
Dirac
operator we introduce 
\bm{c}
H_t  =  \dr(a+\bar A)+t\r_k[\dr(\bar A)].
\eg
This operator is the usual Dirac operator 
for $t=0$ and the infrared
regularised Dirac operator for $t=1$. Now we concentrate on the
evaluation of the zero eigenvalue for $t\ \in\ [0,1]$. We are dealing
with the eigenfunctions $\psi_n(t)$ of $H_t$ with 
\bm{c}
H_t\psi_n(t)  =  E_n(t)\psi_n(t).
\eg
Since $\r_k$ is a compact operator, the normalisability of $\psi_n(t)$
is guaranteed for every $t$. Moreover the Taylor series in $t$ of 
$E_n,\psi_n$ are convergent. 
In the one instanton sector there is one 
eigenvector $\psi_0$ with 
\bm{c}
H_0\psi_0(0)  =  0.
\eg
$E_0(0) =0$. We prove by induction that all derivatives of $E_0$ 
\bm{c}
E_0^{(n)}[t]= \partial_t^n E_0(t)
\eg
are vanishing at $t=0$. This leads to $E_0(t)\ =\ 0$ for $t\in[0,1]$. 
We start with $E_0^{(0)}\ =\ 0$ and assume that $E_0^{(m)}=0$ for all
$m\leq n-1$. 
It follows 
\bm{c}\label{0}
\partial^{m}_t\left[H_t\psi_0(t)\right]_{t=0} =  0\ \ \ \
\forall m\leq n-1
\eg
or 
\bm{ccc}\label{chirality}
\dr(a+\bar A)\psi_0^{(m)}(0) = -m\r_k[\dr(\bar A)] \psi_0^{(m-1)}(0)\ \ \ 
\forall m\leq n-1, & &\di  
\psi_0^{(m)}(t)= \partial_t^m\psi_0(t).
\eg 
As an intermediate result we prove that the
$\psi_0^{(m)}(0)$ are chirality eigenstates with the 
same chirality as $\psi_0(0)$ for all $m\leq n-1$. 
We deduce from (\ref{chirality}) 
\bm{c}\label{chiralpsi}
\gamma_5 \psi_0^{(m)}=  -m\CP\frac{1}{\dr(a+\bar A)} \CP \r_k[\dr(\bar A)] 
\gamma_5\psi_0^{(m-1)}(0)+\gamma_5(1-\CP)\psi_0^{(m)},
\eg 
where $\CP$ is the projector on the space of non-zero modes of 
$\dr(a+\bar A)$ and we have used (see \eq{dirac},(\ref{fermcut}))
\bm{ccc} \label{anticom}
\{\r_k[\dr(\bar A)],\gamma_5\}= \{\dr(a+\bar A),\gamma_5\}=
[\CP,\gamma_5]=0,  & &\di 
\CP  = \left(\dr \frac{1}{\dr^2}\dr\right)(a+\bar A). 
\eg
However $\psi_0^{(0)}(0)$ is a chirality eigenstate, 
$\gamma_5\psi_0(0) =\pm\psi_0(0)$. Furthermore $(1-
\CP)\psi_0^{(m)}(0)$ is proportional to
$\psi_0(0)$. Thus it follows from \eq{chiralpsi} that 
$\psi_0^{(m)}(0)$ has the same chirality as $\psi_0(0)$, 
if $\psi_0^{(m-1)}(0)$ has this property. Starting
iteratively with $m=1$, the claimed chirality properties follow for
all $m\leq n-1$. 

With this result and (\ref{0},\ref{anticom}) we prove $E_0^{(n)}(0)=0$: 
\bg
E_0^{(n)}(0)&  =&\di \!\! 
\partial^n_t\langle\psi_0(t),H_t\psi_0(t)\rangle_{t=0}\\\di
\mbox{eq.~}(\ref{0})\rightarrow   &\di  = &\di \!\!
n\langle\psi_0(t),\r_k[\dr(\bar A)]\partial^{n-1}_t\psi_0(t)\rangle_{t=0}\\\di
\mbox{chirality of }\psi_0^{(n-1)}(0)\rightarrow      &\di  = &\di \!\!
n\langle\gamma_5 \psi_0(0),\r_k[\dr(\bar A)]
\gamma_5\psi_0^{(n-1)}(0)\rangle\\\di 
\mbox{eq.~}(\ref{anticom})\rightarrow   &\di  = &\di \!\!
-n\langle\psi_0(0),\r_k[\dr(\bar A)]\psi_0^{(n-1)}(0)\rangle\\\di 
    &\di  = &\di \!\! 0. 
\eg
Therefore the $E_0^{(n)}(0)$ vanish for all $n\in\N$ 
which leads to $E_0(t)=0,\, t\in [0,1]$. This proves that $\psi_0(t)$ is 
a zero mode for all $t$, in particular for $t=1$. 
           
With these preliminaries we can easily factorise the fermionic 
zero mode
contribution as in the case without regularisation. It follows 
for topologically non-trivial configurations $a+\bar A$
\bm{rl}
Z_{\psi,k}[a,\bar A,\bar\eta,\eta] =&\di \!\!     
\frac{1}{\CN_k}\int\d\bar\chi'\,\d\chi'
\exp\Biggl\{\int_x\bar\chi'\left(\dr(a+\bar A)
+\r_k[\dr(\bar A)]\right)\chi'\\\di  
 &\di    +\int_x\left(\bar\eta'\cdot\chi'+\bar
\chi'\cdot \eta'\right)  \Biggr\}\int\d\bar\chi_0\,\d\chi_0\exp
\int_x\left(\bar\eta
\cdot\chi_0+\bar\chi_0\cdot \eta\right)
\eg
with 
\bm{c}\label{zero_k}
\left(\dr(a+\bar A)+\r_k[\dr(\bar A)]\right)\chi_0 =  0.
\eg

\mysection{Zero mode contribution to leading order of $1/k$}
\label{zeromode}
We recall the expression for $P_k$ (see (\ref{implicit}))
\bm{c}\label{non-local}
P_k[A,\bar A,\psi,\bar\psi] = \int \d\mu_1(\theta)
\prod_{s=1}^{N_f}\int_x\bar\eta_s\cdot\phi_0
\int_x\phi_0^+\cdot\eta_s  +\mbox{h.c.}
\eg 
We shall argue that 
$P_k$ depends only to sub-leading order in $1/k$ on $A,\bar A$. For that 
purpose we concentrate on the zero mode equation \eq{zero_k} with a purely 
topological configuration $a=a_I$. 
In the limit $k\to \infty$ only instantons $a_I(x,\rho)$ 
with width $\rho\sim 1/k$ contribute to \eq{non-local} due to the infrared 
regularisation of the gauge field present in 
$\d\mu_1$ (see appendix~\ref{local}). Note that 
$a_I(x,\rho)=a_I(x/\rho,1)/\rho$ (see e.g.\ \eq{inst},\eq{sing}) and 
$R_k^\psi[\p_x]=R_{k\rho}^\psi[\p_{x/\rho}]/\rho$ (see \eq{fermcut}). 
Hence after multiplying \eq{zero_k} with $\rho\sim1/k\to 0$ and scaling 
$x\to \rho x$ we conclude that 
the fermionic zero mode depends on $A,\bar A$ only 
to sub-leading order. Thus $P_k$ is $A,\bar
A$-independent to leading order and we write 
\bg
P_k[A,\bar A,\psi,\bar\psi] =
P_k[\psi,\bar\psi]
+O(1/k).
\eg  
$P_k$ is non-local. In the limit $k\rightarrow 
\infty$ it is possible to write it as a sum of a local contribution and 
terms which are suppressed with powers of $k^{-1}$. In this limit 
we also calculate the normalisation of $P_k$. 

The measure $\d\mu_1$ 
contains integrations over
collective coordinates. The interesting collective coordinates are the 
centre of the instanton $z$, the width $\rho$ and the global gauge rotations 
$g$. The explicit derivation of 
the $\rho,z$ dependence  of $\d\mu_1$ 
is done in appendix~\ref{local}. Moreover the 
instanton $a_I$ and the fluctuations $a'$ decouple in the limit $k\to\infty$ 
which can be used to effectively remove the gauge fixing term for $a_I$ (see 
derivation of \eq{ratio}). 
Hence $\d\mu_1$ also includes 
a measure $\d g_k$ of local gauge degrees of 
freedom in this limit. Note that the cut-off term for $a_I$ singles out those 
local 
gauge degrees of freedom dependent on large momenta. However this will not 
effect the following arguments. 

We will use well-known results from instanton calculations. For details 
we refer the reader to the literature (\cite{thooft2},\cite{shifman}). 
The normalised fermionic zero mode is given by 
\bm{ccccc}\label{explicitzero}
\phi_{0,\xi}^A(x;z,\rho)   = \frac{\sqrt{2}}{\pi}\frac{\rho}{(
(x-z)^2+
\rho^2)^\frac{3}{2}} u_\xi^A, & &\di  
\sum_A u^A\times \bar u^A    =   \frac{1-\gamma_5}{2}, &  &\di
\|\phi_0\|    =  1   
\eg 
and gauge transformations $g(x)\phi_0(x;z,\rho)$ 
of \eq{explicitzero}, where $g(x)$ could be either $g_k(x)$ 
or a global gauge rotation $g$. With $\rho\sim 1/k\rightarrow 0$ we write 
\bm{c}\label{carefull}
\int_x \bar\eta_s\cdot\phi_0 =
\frac{\sqrt{2}}{\pi} \int_x\frac{\rho}{((x-z)^2+
\rho^2)^\frac{3}{2}}\bar\eta_s(x)\cdot u =
\frac{\sqrt{2}}{\pi} \int_x\frac{\rho}{(x^2+
\rho^2)^\frac{3}{2}}\bar\eta_s(x+z)\cdot u.
\eg 
We are interested in the limit $\rho\rightarrow 0$. Therefore we calculate 
\eq{carefull} in an expansion about $\rho=0$. The coefficient related to the
power $\rho^0$ vanishes. The coefficient proportional to $\rho$ 
is determined by scaling (\ref{carefull}) with $\rho^{-1}$  
\bm{c}\label{order1}
\lim_{\rho\rightarrow 0}\frac{1}{\rho} \frac{\sqrt{2}}{\pi}
\int_{x}\frac{\rho}{(x^2+\rho^2)^\frac{3}{2}}\bar\eta_s(x+z)\cdot u 
   =   \frac{\sqrt{2}}{\pi}\int_{x}\Bigl(\frac{1}{x^2}
\Bigr)^\frac{3}{2}
\bar\eta_s(x+z)\cdot u.
\eg 
The term of order $\rho^2$ is calculated by subtracting (\ref{order1}) 
times $\rho$ from (\ref{carefull}). It follows 
\bm{c}
  \lim_{\rho\rightarrow 0}\frac{1}{\rho} 
\frac{\sqrt{2}}{\pi}
\int_{x}\left[\frac{1}{(x^2+\rho^2)^\frac{3}{2}}-\left(\frac{1}{x^2}
\right)^\frac{3}{2}\right]\bar\eta(x+z)_s\cdot u  =  -2^{5/2}\pi\bar\eta_s(z)
\cdot u
\eg 
The final result for the contribution of the fermionic zero mode 
of an instanton with width $\rho\sim 1/k$ and centre $z$ is 
\bm{c}\label{result}
\int_x\! \bar \eta g^{-1}\!\cdot\phi_0 =
\rho\,\frac{\sqrt{2}}{\pi}\int_x\!
\left(\frac{1}{x^2}\right)^\frac{3}{2}\!\bar\eta_s(x+z)g^{-1}(x+z)\cdot u
-\rho^2 2^{5/2}\pi \bar\eta_s(z) g^{-1}(z)\cdot u+O(\rho^3), 
\eg 
Contributions to $P_k$ dependent on the first (non-local) term on 
the right hand side of (\ref{result}) vanish because of the integration over 
local gauge degrees of freedom $g_k$ present in $\d\mu_1$. 
For the evaluation of the second term on the right hand side of \eq{result} 
we concentrate on the integration over global gauge rotations. We get from 
\eq{non-local} by using \eq{result} 
\bm{c}
P_k[\psi,\bar\psi] =
\!\int\!\d^4 z\!\int\!\d\bar\mu_1(\theta)(2^5\pi^2\rho^4)^{N_f}\!
\int\!\d g\!\prod_{s=1}^{N_f}\Bigl(\bar\eta^s(z)g^{-1}\!\cdot u\Bigr)\Bigl(
\bar u\cdot g\eta^s(z)\Bigr)+\mbox{h.c.}+O(1/k), 
\eg 
where $\d g$ is the measure of global gauge rotations and $\d^4 z$ is the 
measure of the centre of the instanton: 
\bm{ccc}\label{split}
\d\mu_1(\theta)=\d\bar\mu_1(\theta)\,\d g\,\d^4 z, & & \di 
\int\d g=1.  
\eg 
For the evaluation of the $g$-integration in $SU(N)$ we use 
\cite{creutz}
\bm{c}\label{groupint}
\int \d g \prod_{i=1}^{N_f}g^{-1}_{A_i \bar A_i}g^{\ }_{\bar B_i B_i}=
a[N,N_f]
\left(\sum_\sigma\prod_{i=1}^{N_f}
\delta_{A_i B_{\sigma(i)}}\delta_{\bar A_i \bar B_{\sigma(i)}}
+\frac{1}{N}\CO_{A_1\bar B_1\cdots \bar A_{N_f}\bar B_{N_f}}\right), 
\eg 
where $\sigma$ are the permutations of $(1,...,N_f)$. 
The tensor $\CO$ is suppressed with $1/N$ and only consists 
of products of Kronecker deltas $\delta_{A_i B_j}\delta_{\bar A_n\bar B_m}$. 
Both $a[N,N_f]$ and $\CO$ are complicated functions of $N,N_f$. 
With (\ref{explicitzero},\ref{groupint}) we get 
\bm{c}\label{groupint2}
\int \d g \prod_{s=1}^{N_f} \bar\eta_s g\cdot u \bar u\cdot g^{-1}\eta_s  
\stackrel{k\rightarrow\infty}{\longrightarrow} 
a[N,N_f]\det_{s,t}\bar\eta_s^{A_s}\frac{1-\gamma_5}{2}\eta_t^{B_t} 
\left(\delta^{A_s B_t}+\frac{1}{N}\CU^{A_1 B_1\cdots A_{N_f} B_{N_f}}\right). 
\eg 
The tensor $\CU$ is related to $\CO$ and involves only products of 
Kronecker deltas $\delta^{A_i B_j}$. However from now on we 
drop the term dependent on $\CU$. This is a suitable approximation 
within a $1/N$-expansion since it carries the same flavor structure 
as the leading term but is supressed with $1/N$. Note however that this is 
done more for the sake of convenience and the 
tensor structure can be added without changing the 
conclusions of the present paper. Moreover even so tedious the calculation of 
$\CU$ is straightforward. This leads to 
\bm{c}
P_k[\psi,\bar\psi] 
  \stackrel{k\rightarrow\infty,\ N\gg 1}{\longrightarrow}  
\int\d^4 z \int\d\bar\mu_1(\theta)\,
(2^5\pi^2\rho^4)^{N_f}
a[N,N_f]\det_{s,t}\bar\eta_s^A(z)\frac{1-\gamma_5}{2}\eta^A_t(z)+\mbox{h.c.}
\eg
Now we are able to give a final expression for $P_k$ 
\bm{c}\label{finresult}
P_k[\psi,\bar\psi] =  
\int_z\Delta[k,\theta]\det_{s,t}\bar\eta_s(z)
\frac{1-\gamma_5}{2}\eta_t(z)+O(\Delta[k,\theta]/k) 
\eg
with  
\bm{c}\label{scaling}
\Delta[k,\theta]= \int
\d\bar\mu_1(\theta)\,(2^5\pi^2\rho^4)^{N_f}
a[N,N_f]\ \sim  k^{-5N_f+4}. 
\eg 
The gauge field cut-off ensures the finiteness of 
$\Delta[k,\theta]$. Then the $k$-dependence follows by dimensional arguments.

\mysection{Properties of the gauge field regularisation}\label{local}
In this section we examine the properties of the cut-off term for the gauge 
field in the 1 instanton sector. A general instanton is given by  
\bm{c}\label{inst}
A_{I,\mu}^a(x;z,\rho)    =    \eta_{\mu\nu}^a 
\frac{(x-z)_\nu}{(x-z)^2+\rho^2}
\label{instanton}\eg
and global gauge rotations of $A_{I,\mu}^a(x;z,\rho)$. 
Here $\eta^a_{\mu\nu}$ are the 't Hooft symbols \cite{thooft2}. 
In order to stay in contact with \cite{ellwanger} 
we first discuss this approach where the background field is missing. 
In this case
 the field $a$ consists on both, the instanton (\ref{instanton}) and the
  fluctuations about the instanton. The configurations 
(\ref{instanton}) are 
not square-integrable because of their infrared behaviour.  To see this,  
let us recall the cut-off 
term for the gauge field (see (\ref{gaugecut}) and \eq{limits}) 
\bm{ccc}\label{recall}
 \Delta_k S_A[a,0] = \frac{1}{2}\int_x a\cdot  
R^{A}_{k}[D_T(0)]\cdot a, & &\di R_{k,\mu\nu}^{A,ab}[D_T(0)] 
\stackrel{\frac{\|D_T(0)\|}{k^2}\rightarrow 0}{\longrightarrow}   k^2
\delta^{ab}\delta_{\mu\nu}.
\eg  
If a configuration $a$ is ultraviolet finite but is not 
square-integrable due to infrared divergences, then 
$\Delta_k S_A[a,0]$ diverges. These configuration have zero 
measure in the path integral, since   
\bm{c}
\exp\{-\Delta_k S_A[a,0]\}   =    0.
\eg 
We conclude that only configurations which decrease faster than
$1/x^2$ can contribute to
the infrared regularised path integral. This reflects the fact 
that within this particular approach \cite{ellwanger} the cut-off term 
introduces trivial (infrared) boundary-conditions. Thus the infrared
cut-off term (without background field dependence) 
introduces a constraint on the class of gauge fixings, i.e.\ allowing 
only for those compatible with trivial infrared behaviour. 
Instantons in the singular
gauge fulfill this condition (see for example \cite{shifman}). 
They are given by 
\bm{c}\label{sing}
a_{I,\mu}^a(x;z,\rho)=\eta_{\mu\nu}^a
\frac{(x-z)_\nu}{(x-z)^2}
\frac{\rho^2}{(x-z)^2+\rho^2}.
\eg
These configurations are square-integrable, so 
$\Delta_k S_A[a,\bar A]$ is finite. We write
explicitly for an instanton $a_I(x,z,\rho)$ with centre $z$ and width
$\rho$ 
\bm{c}
\Delta_k S_A[a_I(x,z,\rho),0] =  
\frac{1}{2}\int_{\tilde x} a_{I}(\tilde x,0,1)\cdot 
R_{k\rho}^{A}[-\partial_{\tilde x}^2]\cdot a_{I}(\tilde x,0,1),
\eg 
where we have used the translation invariance of the cut-off term for $D_T(0)$ 
and have changed the variable $x$ to $\tilde x= (x+z)/\rho$. Using the limit 
in \eq{recall} we get 
\bm{c}\label{explicit}
\exp\left\{-\Delta_k S_A[a_I(x,z,\rho),0]\right\}
\stackrel{k\rho\gg 1}{\longrightarrow} \exp\left\{-(k\rho)^2\frac{1}{2}
\int_{\tilde x} a_{I}(\tilde x,0,1)\cdot a_{I}(\tilde x,0,1)\right\} 
\sim e^{-\#(k\rho)^2}.
\eg 
Hence in the limit $k\rightarrow\infty$ only instantons with width 
$\rho\sim k$ contribute. 
This result extends easily to the more general case with non-vanishing 
background 
field $\bar A$. Moreover the constraint on the class of gauge fixings 
is related entirely to the introduction of trivial infrared boundary 
conditions by choosing $\bar A=0$. Following 
the background field approach to instantons \cite{thooft1,thooft2} one 
chooses the background field $\bar A$ as the configuration \eq{instanton}. 
In this case the field $a$ consists on fluctuations about the instanton which 
are square-integrable by definition (after a complete gauge fixing). 
This leads immediately to \eq{explicit}.

\end{appendix}

\end{document}